\documentclass[journal]{IEEEtran}

\usepackage{lineno,hyperref}
\usepackage{bm}
\usepackage{mathrsfs}
\usepackage{amsmath}
\usepackage{amssymb}
\usepackage{graphicx}
\usepackage{subfigure}
\usepackage{epstopdf}
\usepackage{xcolor}
\usepackage{wasysym}
\usepackage{footnote}
\usepackage{amssymb}
\usepackage{overpic}
\modulolinenumbers[5]

\ifCLASSINFOpdf

\else

\fi

\begin{document}

\title{Study of General Robust Subband Adaptive Filtering}

\author{Yi~Yu,~\IEEEmembership{Member,~IEEE},
    ~Hongsen He,~\IEEEmembership{Member,~IEEE},
    ~Rodrigo C. de Lamare,~\IEEEmembership{Senior Member,~IEEE},\\
    ~Badong Chen,~\IEEEmembership{Senior Member,~IEEE}

    \thanks{This work was supported in part by the National Natural Science Foundation of China (Nos. 61901400, 62071399), Natural Science Foundation of Sichuan (No. 2022NSFSC0542), Sichuan Science and Technology Program (No. 2021YFG0253), and Doctoral Research Fund of Southwest University of Science and Technology in China (No. 19zx7122).  (Corresponding author: Hongsen He). }

    \thanks{Y. Yu and H. He are with School of Information Engineering, Robot Technology Used for Special Environment Key Laboratory of Sichuan Province, Southwest University of Science and Technology, Mianyang, 621010, China (e-mail: yuyi\_xyuan@163.com, hongsenhe@gmail.com).}


    \thanks{R. C. de Lamare is with CETUC, PUC-Rio, Rio de Janeiro 22451-900, Brazil. (e-mail: delamare@cetuc.puc-rio.br).}

    \thanks{B. Chen is with Institute of Artificial Intelligence and Robotics, Xi'an Jiaotong University, Xi'an, 710049, Shaanxi Province, China. (e-mail: chenbd@mail.xjtu.edu.cn).}


}


\maketitle

\begin{abstract}
In this paper, we propose a general robust subband adaptive filtering (GR-SAF) scheme against impulsive noise by minimizing the mean square deviation under the random-walk model with individual weight uncertainty. Specifically, by choosing different scaling factors such as from the M-estimate and maximum correntropy robust criteria in the GR-SAF scheme, we can easily obtain different GR-SAF algorithms. Importantly, the proposed GR-SAF algorithm can be reduced to a variable regularization robust normalized SAF algorithm, thus having fast convergence rate and low steady-state error. Simulations in the contexts of system identification with impulsive noise and echo cancellation with double-talk have verified that the proposed GR-SAF algorithms outperforms its counterparts.
\end{abstract}


\IEEEpeerreviewmaketitle
\section{Introduction}
\IEEEPARstart{N}{owadays}, adaptive filters have been applied in a
variety of applications such as system identification, echo
cancellation, active noise control, and acoustic feedback
cancellation~\cite{sayed2003fundamentals,lee2009subband,paleologu2010efficient,yang2011proportionate,
pradhan2017acoustic,yin2021robust},\cite{jidf,spa,intadap,mbdf,jio,jiols,jiomimo,sjidf,ccmmwf,tds,mfdf,l1stap,mberdf,jio_lcmv,locsme,smtvb,ccmrls,dce,itic,jiostap,aifir,ccmmimo,vsscmv,bfidd,mbsic,wlmwf,bbprec,okspme,rdrcb,smce,armo,wljio,saap,vfap,saalt,mcg,sintprec,stmfdf,1bitidd,jpais,did,rrmber,memd,jiodf,baplnc,als,vssccm,doaalrd,jidfecho,dcg,rccm,ccmavf,mberrr,damdc,smjio,saabf,arh,lsomp,jrpaalt,smccm,vssccm2,vffccm,sor,aaidd,lrcc,kaesprit,lcdcd,smbeam,ccmjio,wlccm,dlmme,listmtc,smcg,mfsic,cqabd,rmmse,rsthp,dmsmtc,dynovs,dqalms,detmtc,1bitce,mwc,dlmm,rsbd,rdcoprime,rdlms,lbal,wlbd,rrser},
. In adaptive filtering, the least mean square (LMS) and normalized
LMS (NLMS) algorithms are widely investigated due to the low
computational complexity of $\mathcal{O}(M)$. Also, unlike the LMS,
the stability of the~NLMS is independent of the maximum eigenvalue
of the input autocorrelation matrix. Nevertheless, for highly
correlated input signals (also called colored input signals), these
two algorithms converge slowly.

For such input signals, both the families of affine projection (AP)~\cite{ozeki1984adaptive,choi2007adaptive} and recursive least squares (RLS)~\cite{sayed2003fundamentals,paleologu2008robust} have fast convergence rate, but they require much higher computational complexity especially for large~$M$. Although the low-complexity AP and RLS variants were also studied~\cite{sayed2003fundamentals,yang2018comparative,zakharov2008low}, some of them would face the numerical instability problem due to round-off errors. On the other hand, thanks to the inherent decorrelation property of subband adaptive filtering (SAF) on input signals, its multiband structure has received much attention~\cite{lee2009subband}. Specifically, the input signals are divided into multiple subbands signals through the analysis filters, and in each subband the decimated input sequences are approximately white while updating the weight vector. Therefore, the normalized SAF (NSAF) algorithm proposed by Lee and Gan~\cite{lee2004improving} achieves much faster convergence rate than the NLMS algorithm in the case of colored input signals. It is noteworthy that the NSAF algorithm has almost comparable computational complexity to the NLMS algorithm with~$\mathcal{O}(M)$. As a result, the NSAF algorithm is promising for applications of large~$M$, like in the echo cancellation. For the time-delay problem recovering the signal from the decimated sequences through the synthesis filter bank, the delayless variants of NSAF were proposed in~\cite{lee2007delayless}, which compute the output of adaptive system in an auxiliary loop.

Regrettably, the above algorithms suffer from the performance degradation in the non-Gaussian noise scenarios with impulsive samples. The impulsive noise frequently happens in echo cancellation, underwater acoustics, audio processing, and communications~\cite{nikias1995signal,zimmermann2002analysis,georgiou1999alpha},~etc. In impulsive noise scenarios, thus studying robust
adaptive filtering algorithms is also vital and in the literature a variety of robust algorithms have also been reported~\cite{zhou2011new,huang2017maximum,chen2016generalized,al2016robust,huang2019norm,pogula2019robust,rakesh2019modified,radhika2021proportionate}. There are some typical examples, i.e., M-estimate based algorithm~\cite{zhou2011new}, maximum Versoria criterion based algorithm~\cite{huang2017maximum,radhika2021proportionate}, maximum correntropy criterion (MCC) based algorithm~\cite{chen2016generalized}, and Lorentzian norm based algorithm~\cite{huang2019norm}, to name a few. Likewise, to obtain the robustness of SAFs against impulsive noise, the sign SAF (SSAF) algorithm was proposed in~\cite{ni2010signsubband}. In order to take full advantage of the decorrelation of SAF dealing with input signals, reference~\cite{yu2016novel} developed the individual-weighting-factors based SSAF (IWF-SSAF) algorithm, which has shown faster convergence rate than the SSAF algorithm. Also, Ni~\emph{et al} extended the AP concept to the SSAF~\cite{ni2014two}, the resulting AP-SSAF has better convergence. However, the SSAF-type still has relatively slow convergence, since the weights' update does not exploit the magnitude information of the subband error signals no matter how impulsive noises appear or not. To overcome this shortcoming, a class of step-size scaling factor based NSAF algorithms were proposed~\cite{hur2016variable, huang2017combined}, which diminish sharply the step-size when impulsive samples appear, thus obtaining better convergence behavior as compared to the IWF-SSAF algorithm. Roughly speaking, these algorithms originate from the combination of specified robust cost functions and the NSAF update. Furthermore, by defining the M-estimate based robust cost function, the AP M-estimate SAF (APM-SAF) algorithm was proposed in~\cite{zheng2016affine} with better convergence for the colored inputs in impulsive noises, at the expense of high computational complexity due to the AP principle.

On the other hand, the family of SAF algorithms exists a trade-off between convergence rate and steady-state error on choosing the step-size. In response to this problem, various variable step-size (VSS)~\cite{ni2010variable,jeong2012variable} and variable regularization parameter~\cite{ni2010normalised,shin2018adaptive} schemes were designed for the NSAF algorithm, to obtain fast convergence and low steady-state error simultaneously in the Gaussian noise. In the existing literature~\cite{kim2013sign,yoo2014band,kim2017delayless,YU2021107806}, authors presented different VSS algorithms for both SSAF and IWF-SSAF to further improve their performance in the presence of impulsive noises, and in~\cite{YU2021107806} authors also considered the sparsity of the unknown system about the performance improvement. Nonetheless, they generally have no tracking capability when the unknown systems change abruptly, due to the fact that the time-varying step-size is always decreasing monotonically to ensure the robustness against impulsive noises. In~\cite{yu2019m}, the authors analyzed the performance of the M-estimate based NSAF (M-NSAF) algorithm in impulsive noises, and then designed a VSS scheme for the M-NSAF algorithm. The resulting VSS-M-NSAF algorithm achieves fast convergence and low steady-state error. However, this VSS has no generality for robust NSAF algorithms with scaling factors.

In this paper, the main contributions are as follows.

1) By reviewing the M-NSAF algorithm, we present a general scheme that employs a robust weights update rule with scaling factors. Then by introducing the concept of the probability update, we derive a general robust SAF (GR-SAF) algorithm against impulsive noise that minimizes the mean square deviation (MSD) under the random-walk model with individual uncertainty for every filter weight.

2) Importantly, by setting the scaling factors from a specified robust criteria, we can straightforwardly obtain the corresponding GR-SAF algorithm.

3) The proposed GR-SAF algorithm can also be interpreted as a robust NSAF algorithm with variable regularization, thus it exhibits fast convergence and low steady-state error in the impulsive noise.

The remaining part of this paper is organized as follows. Section~II reviews M-NSAF adaptive filtering for the delayless SAF structure and states the problem. In Section~III, we present the GR-SAF scheme and give its implementation whereas Section IV presents an analysis of several aspects of the GR-SAF algorithm. In Section~V, extensive simulations are shown. Conclusions are drawn in Section~VI.

\section{M-NSAF Adaptive Filtering and Problem Statement}
\begin{figure*}[htb]
    \centering
    \includegraphics[scale=0.55] {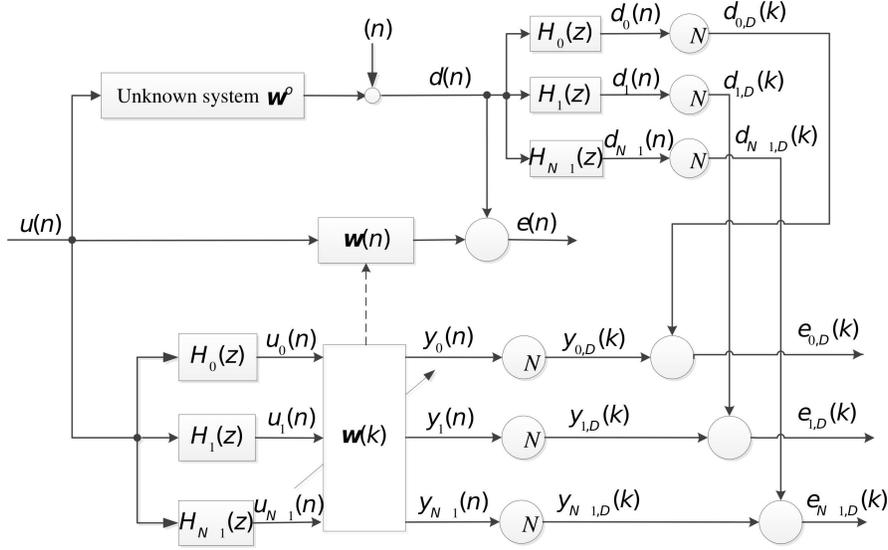}
    \hspace{2cm}\caption{Delayless multiband-structured SAF.}
    \label{Fig1}
\end{figure*}
Let us consider a system identification problem whose input-output data pair $\{u(n),d(n)\}$ at every time index $n$ obeys the following relation:
\begin{align}
\label{001}
d(n) = \bm u^\text{T}(n) \bm w^o+\nu(n),
\end{align}
where $(\cdot)^\text{T}$ denotes the transpose, and $\bm u(n)=[u(n), u(n-1),...,u(n-M+1)]^\text{T}$ is the $M$-dimensional input vector independent of the additive noise $\nu(n)$. The impulse response of the unknown system, denoted as the $M\times1$ vector~$\bm w^o$, needs to be estimated by an adaptive filter $\bm w(n)$ with the same input signal $u(n)$. For this purpose, Fig.~\ref{Fig1} illustrates a delayless multiband-structured SAF~\cite{lee2007delayless,kim2017delayless}, where $N$ is the number of subbands. Both signals $d(n)$ and $u(n)$ are decomposed by the analysis filters $H_i(z),\;i=0,...,N-1$, to yield the band-dependent signals $d_i(n)$ and $u_i(n)$, respectively. Giving the subband input signal~$u_i(n)$, we obtain the subband output signal~$y_i(n)$ of the adaptive filter. Then, $d_i(n)$ and $y_i(n)$ are critically sub-sampled to a lower rate, respectively resulting in the decimated subband signals $d_{i,D}(k)=d_i(kN)$ and $y_{i,D}(k)=\bm u^\text{T}_i(k) \bm w(k-1)$, where $\bm u_i(k)=[u_i(kN),u_i(kN-1),...,u_i(kN-M+1)]^\text{T}$. In the decimated sequences, the weight vector $\bm w(k)$ is the estimate of $\bm w^o$ and controlled by the decimated subband error signals given by
\begin{equation}
\label{002}
\begin{array}{rcl}
\begin{aligned}
e_{i,D}(k) &= d_{i,D}(k) - y_{i,D}(k)\\
&=d_{i,D}(k) - \bm u^\text{T}_i(k) \bm w(k-1),
\end{aligned}
\end{array}
\end{equation}
for $i=0,...,N-1$.

In some practical cases such as the EC, we might focus on the output error $e(n)$ in the original sequences. As such, we copy $\bm w(k)$ to $\bm w(n)$ for every $N$ input samples (i.e., when $n=kN$), and compute the output error $e(n) = d(n)-\bm u^\text{T}(n) \bm w(n)$ in an auxiliary loop. The additive noise may include not only the Gaussian noise but also impulsive noise. To obtain robustness against impulsive noise, therefore the M-NSAF algorithm for updating the weight vector is described as~\cite{yu2019m}
\begin{equation}
\label{003}
\begin{array}{rcl}
\begin{aligned}
\bm w(k) = \bm w(k-1) + \mu \sum \limits_{i=0}^{N-1}\frac{\varphi'(e_{i,D}(k))\bm u_i(k)}{||\bm u_i(k)||_2^2},\\
\end{aligned}
\end{array}
\end{equation}
where $||\cdot||_2$ denotes the $l_2$-norm of a vector, and $\mu>0$ is the step-size. The score function is given by
\begin{equation}
\label{004}
\varphi'(x)\triangleq \frac{\partial\varphi(x)}{\partial x}=\left\{ \begin{aligned}
&x, \text{ if } |x| < \xi\\
&0,\text{ if } |x| \geq \xi,
\end{aligned} \right.
\end{equation}
where $\varphi(x)$ is the modified Huber (MH) function
\begin{equation}
\label{005}
\varphi(x)=\left\{ \begin{aligned}
& x^2/2, \text{ if } |x| < \xi\\
&\xi^2/2, \text{ if } |x| \geq \xi,
\end{aligned} \right.
\end{equation}
and $\xi$ is a threshold parameter.

Equations~\eqref{003} and~\eqref{004} clearly show, when $|e_{i,D}(k)|<\xi$ holds, the M-NSAF algorithm will reduce to the standard NSAF algorithm. Once the values of $|e_{i,D}(k)|$ are larger than $\xi$ (may be due to the occurrence of outliers or impulsive noise), the M-NSAF algorithm will freeze the adaptation of the weight vector to guarantee the stability of the algorithm. That is to say, the threshold $\xi$ controls the capability of suppressing impulsive noise. At each subband, the threshold is chosen as $\xi_i =\kappa \hat{\sigma}_{e,i}(k)$~\cite{zhou2011new}, where $\sigma_{e,i}^2(k)$ denotes the estimated variance of 'free-impulsive' subband error and is calculated by the recursion
\begin{equation}
\label{006}
\begin{array}{rcl}
\begin{aligned}
\hat{\sigma}_{e,i}^2(k) = \theta \hat{\sigma}_{e,i}^2(k-1) + c_\sigma(1-\theta) \text{med}(\bm a_{e,i}(k)).
\end{aligned}
\end{array}
\end{equation}
Here, we choose the typical value~2.576 for $\kappa$, which means the 99\% confidence of rejecting $e_{i,D}(k)$ participated in the adaptation~\eqref{003} when $|e_{i,D}(k)| \geq \xi_i$, under the Gaussian assumption of $e_{i,D}(k)$ except when being contaminated accidentally by impulsive noise. In \eqref{006}, $0 \ll \theta <1$ is the weighting factor which is often chosen by $\theta=1-N/(\tau M)$ with $\tau \geq1$ (while $\theta=0$ at the initial iteration $k=0$), and $c_\sigma=1.483(1+5/(N_w-1))$ is the correction factor. The median operator~$\text{med}(\cdot)$ is to discard the outliers over the sliding data window $\bm a_{e,i}(k)=[e_{i,D}^2(k),e_{i,D}^2(k-1),...,e_{i,D}^2(k-N_w+1)]$ with $N_w$-length, where the value of $N_w$ would be properly increased for the long-lasting occurrence of impulsive noises like the double-talk.

However, the main problem for the M-NSAF algorithm is the slow convergence and large steady-state error which depends on how to choose the step size $\mu$. In this paper, we aim at solving this problem by developing an algorithm that has fast convergence and low steady-state error.
\section{Proposed GR-SAF Scheme and Algorithms}
In this section, we present the proposed GR-SAF framework and the GR-SAF algorithm based on the probability update mechanism, which is not only robust against impulsive noise but also has fast convergence and low steady-state estimation error.
\subsection{Derivation of Update}
To begin with, we can rearrange~\eqref{003} as
\begin{equation}
\label{007}
\begin{array}{rcl}
\begin{aligned}
\bm w(k) = \bm w(k-1) + \sum \limits_{i=0}^{N-1} q(e_{i,D}(k)) \bm g_i(k) e_{i,D}(k),
\end{aligned}
\end{array}
\end{equation}
where
\begin{equation}
\label{008}
q(x)\triangleq \frac{\varphi'(x)}{x}=\left\{ \begin{aligned}
&1, \text{ if } |x| < \xi\\
&0,\text{ if } |x| \geq \xi,
\end{aligned} \right.
\end{equation}
is called the scaling parameter, and $\bm g_i(k)$ is a free-design vector. Note that,~\eqref{003} and~\eqref{007} are equivalent when $\bm g_i(k)=\mu \frac{\bm u_i(k)}{||\bm u_i(k)||_2^2}$.

To move forward, we assume that the unknown vector $\bm w^o$ follows from the random-walk model~\cite{sayed2003fundamentals} in the decimated domain,
\begin{equation}
\label{009}
\begin{array}{rcl}
\begin{aligned}
\bm w^o(k) = \bm w^o(k-1) + \bm c(k),\\
\end{aligned}
\end{array}
\end{equation}
where $\bm c(k)$ indicates the random disturbance vector with zero mean and covariance matrix $\bm C_w = \text{diag}\{\sigma_{w1}^2,\sigma_{w2}^2..., \sigma_{wM}^2\}$. This implies that each unknown coefficient has an independent level of uncertainty. Consequently, $\bm w^o$ is assumed to be slowly time-varying in the original domain. Accordingly, the decimated subband desired signals can be formulated as
\begin{equation}
\label{010}
\begin{array}{rcl}
\begin{aligned}
d_{i,D}(k) = \bm u^\text{T}_i(k) \bm w^o(k)+ \nu_{i,D}(k),
\end{aligned}
\end{array}
\end{equation}
and the decimated subband inputs and noises are given by respectively
\begin{equation}
\label{011}
\begin{array}{rcl}
\begin{aligned}
\bm u_i(k) & =[\bm u(kN),...\bm u(kN-J+1)] \bm h_i\\
\nu_{i,D}(k)&=\bm h_i^\text{T} [\nu(kN),...,\nu(kN-J+1)]^\text{T},
\end{aligned}
\end{array}
\end{equation}
where $\bm h_i$ is the impulse response of the $i$-th analysis filter with length of $J$.

Subtracting both sides of~\eqref{007} from~\eqref{009}, we have
\begin{equation}
\label{012}
\begin{array}{rcl}
\begin{aligned}
\widetilde{\bm w}(k) = \widetilde{\bm w}(k-1) - \sum \limits_{i=0}^{N-1} q(e_{i,D}(k)) \bm g_i(k) e_{i,D}(k) +\bm c(k),\\
\end{aligned}
\end{array}
\end{equation}
where $\widetilde{\bm w}(k) \triangleq \bm w^o(k) - \bm w(k)$ denotes the weight-error vector.

As can be seen in~\eqref{008}, when~$|e_{i,D}(k)| < \xi_i$ is true, we will have~$q_i(k)=1$, which means that at the iteration~$k$ the $i$-th subband's signals participate in the update of the filter weights. Otherwise, we have~$q_i(k)=0$ (usually that is caused by the impulsive noise), which signifies that at the iteration~$k$ the $i$-th subband's signals have no contribution on updating the filter weights. Hence, we can define a parameter $P_{e,i}(k)\in [0, 1]$ modeling the probability of the $i$-th subband's signals used to update the filter weights at iteration $k$, i.e.,

\begin{equation}
\label{013}
\begin{array}{rcl}
\begin{aligned}
P_{e,i}(k) = P\{|e_{i,D}(k)| < \xi_i\}.
\end{aligned}
\end{array}
\end{equation}
Then, with $P_{e,i}(k)$ instead of $q_i(k)$ we rewrite~\eqref{012} as,
\begin{equation}
\label{014}
\begin{array}{rcl}
\begin{aligned}
\widetilde{\bm w}(k) = \widetilde{\bm w}(k-1) - \sum \limits_{i=0}^{N-1} P_{e,i}(k) \bm g_i(k) \hat{e}_{i,D}(k) +\bm c(k).
\end{aligned}
\end{array}
\end{equation}
Note that, $\hat{e}_{i,D}(k)$ in~\eqref{014} does not include the impulsive noise, as the effect of impulsive noise has been transferred to $P_{e,i}(k)$. Although~\eqref{012} and~\eqref{014} are distinct, it can be shown that their expected values are equivalent. Since our target is to study the average behavior of the algorithm,~\eqref{014} fits well the needs\footnote{A similar treatment was first introduced in the analysis of set-membership filtering algorithm~\cite{diniz2003set}.}. By combining~\eqref{009} and~\eqref{010}, $\hat{e}_{i,D}(k)$ further becomes
\begin{equation}
\label{015}
\begin{array}{rcl}
\begin{aligned}
\hat{e}_{i,D}(k) &= \bm u^\text{T}_i(k) \widetilde{\bm w}(k-1) + \bm u^\text{T}_i(k) \bm c(k) + \nu_{i,D}(k).
\end{aligned}
\end{array}
\end{equation}

After inserting~\eqref{015} into~\eqref{014}, we enforce the operator $\bm \Phi(k) \triangleq \widetilde{\bm w}(k) \widetilde{\bm w}^\text{T}(k)$ on both sides of the equation to establish
\vspace{-0.5cm}
\begin{equation}
\label{016}
\begin{array}{rcl}
\begin{aligned}
\bm \Phi&(k) = \bm \Phi(k-1)  - \sum \limits_{i=0}^{N-1} P_{e,i}(k) \bm \Phi(k-1) \bm u_i(k)\bm g^\text{T}_i(k)  \\
& - \sum \limits_{i=0}^{N-1} P_{e,i}(k) \bm g_i(k) \bm u^\text{T}_i(k) \bm \Phi(k-1)    \\
& + \sum \limits_{i=0}^{N-1} P_{e,i}^2(k) \bm g_i(k)\bm g^\text{T}_i(k)  \left[  \bm u^\text{T}_i(k) \bm \Phi(k-1) \bm u_i(k) + \right. \\
& \left. \bm u^\text{T}_i(k) \bm c(k)\bm c^\text{T}(k) \bm u_i(k) + \nu_{i,D}^2(k) \right] + \bm c(k)\bm c^\text{T}(k) + \bm t_{c}, \\
\end{aligned}
\end{array}
\end{equation}
where $\bm t_{c}$ stands for the cross terms on $\bm u_i(k)$, $\nu_{i,D}(k)$, and $\widetilde{\bm w}(k)$. In the above relation we have also used the orthogonal approximation of decimated input vectors of different subbands~\cite{lee2009subband}. Now, we require the known \emph{independence assumption} in adaptive filtering that $\widetilde{\bm w}(k-1)$ is statistically independent of $\bm u_i(k)$ for $i=0,...,N-1$~\cite{jeong2016mean,zhang2019mean,yin2011stochastic}. With this, we can take the expectations conditioned on $\bm u_i(k)$ on both sides of~\eqref{016} as follows:
\vspace{-0.5cm}
\begin{equation}
\label{017}
\begin{array}{rcl}
\begin{aligned}
\bar{\bm \Phi}&(k) = \bar{\bm \Phi}(k-1)  - \sum \limits_{i=0}^{N-1} P_{e,i}(k) \bar{\bm \Phi}(k-1) \bm u_i(k)\bm g^\text{T}_i(k)   \\
& - \sum \limits_{i=0}^{N-1} P_{e,i}(k) \bm g_i(k) \bm u^\text{T}_i(k) \bar{\bm \Phi}(k-1) \\
& + \sum \limits_{i=0}^{N-1} P_{e,i}^2(k) \bm g_i(k)\bm g^\text{T}_i(k) \left[  \bm u^\text{T}_i(k) \bar{\bm \Phi}(k-1) \bm u_i(k) + \right. \\
&\left. \bm u^\text{T}_i(k) \bm C_w \bm u_i(k) + \sigma_{\nu,i}^2 \right] + \bm C_w,\\
\end{aligned}
\end{array}
\end{equation}
where $\bar{\bm \Phi}(k)=\text{E}\{\bm \Phi(k)\}$ denotes the covariance matrix of $\widetilde{\bm w}(k)$, $\text{E}\{\bm t_{c}\}=0$, and $\sigma_{\nu,i}^2 =\text{E}\{\nu_{i,D}^2(k)\}= ||\bm h_i||_2^2 \sigma_\nu^2$, with $\text{E}\{\cdot\}$ being the mathematical expectation. Taking the traces of all the terms in~\eqref{017}, defined as $J(k)=\text{Tr} \{\bar{\bm \Phi}(k) \}$ that is also called the MSD, we have
\begin{equation}
\label{018}
\begin{array}{rcl}
\begin{aligned}
J(k) =& J(k-1)  - 2\sum \limits_{i=0}^{N-1} P_{e,i}(k) \bm g^\text{T}_i(k) \bar{\bm \Phi}(k-1) \bm u_i(k) \\
& + \sum \limits_{i=0}^{N-1} P_{e,i}^2(k) ||\bm g_i(k)||_2^2 \left[ \bm u^\text{T}_i(k) \bar{\bm \Phi}(k-1) \bm u_i(k) + \right. \\
& \left. \bm u^\text{T}_i(k) \bm C_w \bm u_i(k) + \sigma_{\nu,i}^2\right]  + \text{Tr}\{\bm C_w\}.\\
\end{aligned}
\end{array}
\end{equation}

Then, by setting the derivative of $J(k)$ with respect to $g_i(k)$, $i=0,...,N-1$, to be zero, we obtain
\begin{equation}
\label{019}
\begin{array}{rcl}
\begin{aligned}
&\bm g_i(k) = \frac{1}{P_{e,i}(k)} \times \\
&\frac{\bar{\bm \Phi}(k-1) \bm u_i(k) }{\left[ \bm u^\text{T}_i(k) \bar{\bm \Phi}(k-1) \bm u_i(k) + \bm u^\text{T}_i(k) \bm C_w \bm u_i(k) + \sigma_{\nu,i}^2\right] },
\end{aligned}
\end{array}
\end{equation}
which arrives at the minimization of MSD at each iteration~$k$. Recalling that the probability of occurring impulsive noises is often small, which makes $P_{e,i}(k)<1$ but $P_{e,i}(k)$ close to 1. Accordingly, an efficient and practical scheme for $\bm g_i(k)$ can be obtained from~\eqref{019}:
\begin{equation}
\label{020}
\begin{array}{rcl}
\begin{aligned}
\bm g_i(k) = \frac{\bar{\bm \Phi}(k-1) \bm u_i(k) }{\bm u^\text{T}_i(k) \bar{\bm \Phi}(k-1) \bm u_i(k) + \bm u^\text{T}_i(k) \bm C_w \bm u_i(k) + \sigma_{\nu,i}^2}.
\end{aligned}
\end{array}
\end{equation}
By substituting~\eqref{020} into \eqref{017}, it is established that
\begin{equation}
\label{021}
\begin{array}{rcl}
\begin{aligned}
\bar{\bm \Phi}(k) =&  \bar{\bm \Phi}(k-1) - \sum \limits_{i=0}^{N-1} \left[ 2P_{e,i}(k) - P_{e,i}^2(k))\right]  \times\\
&\bm g_i(k) \bm u^\text{T}_i(k) \bar{\bm \Phi}(k-1) + \bm C_w.
\end{aligned}
\end{array}
\end{equation}
Recalling the internal relation between $P_{e,i}(k)$ and $q(e_{i,D}(k))$, we develop the applicable update rule for $\bar{\bm \Phi}(k)$:
\begin{equation}
\label{022}
\begin{array}{rcl}
\begin{aligned}
\bar{\bm \Phi}(k) = \bar{\bm \Phi}(k-1) - \sum \limits_{i=0}^{N-1} \rho_i(k) \bm g_i(k) \bm u^\text{T}_i(k) \bar{\bm \Phi}(k-1) + \bm C_w,
\end{aligned}
\end{array}
\end{equation}
where $\rho_i(k)=2q(e_{i,D}(k)) - q^2(e_{i,D}(k)$.

\subsection{Practical Considerations}
Equations~\eqref{007},~\eqref{020}, and \eqref{022} constitute the update recursions of the proposed GR-SAF algorithm; however about its implementation, we have to take into account the following aspects.

1) The covariance matrix update~\eqref{022} provides the optimal performance of the GR-SAF algorithm in theory, since it acts like the robust extension of Kalman filtering framework in the subband domain, but it leads also to two problems. First, similar to the Kalman filtering~\cite{ardalan1987fixed,hsu1982square}, the covariance matrix update~\eqref{022} can be numerically unstable when implemented in finite precision, since it is computed as a difference of two positive semidefinite matrices. Numerical accuracy is reduced in every iteration. Moreover, numerical deterioration is also associated with high dimensionality of $\bm g_i(k)$ and $\bar{\bm \Phi}(k)$ on $\bm u_i(k)$ and with the accumulated effects of round-off error. Second, the computation of~$\bar{\bm \Phi}(k)$ requires the complexity of $\mathcal{O}(M^2)$ which is high for large~$M$. To overcome these two points, we always limit $\bar{\bm \Phi}(k)$ to be a diagonal matrix. By doing so, the proposed GR-SAF algorithm can be implemented in the vector form, which reduces to~the complexity of~$\mathcal{O}(M)$ (see subsection~IV.~D). Herein, $\bar{\bm \Phi}(k)$ can be easily initialized by~$\bar{\bm \Phi}(0)=\frac{\epsilon}{M} \bm I_M$, where $\epsilon\geq1$.

2) Both~\eqref{020} and \eqref{022} require knowing $\bm C_w = \text{diag}\{\sigma_{w1}^2,\sigma_{w2}^2..., \sigma_{wM}^2\}$ beforehand which controls a trade-off problem between tracking capability and steady-state behavior of the algorithm. From~\eqref{009}, we have the following relation for the $m$-th diagonal entry:
\begin{equation}
\label{023x}
\begin{array}{rcl}
\begin{aligned}
\sigma_{wm}^2 = \text{E}\{[w_m^o(k)- w_m^o(k-1)]^2\}, m=1,...,M-1.
\end{aligned}
\end{array}
\end{equation}
Accordingly, for estimating them the following practical recursion is used:
\begin{equation}
\label{023}
\begin{array}{rcl}
\begin{aligned}
\hat{\sigma}_{wm}^2(k) = \gamma \hat{\sigma}_{wm}^2(k-1) + (1-\gamma)[w_m(k)-w_m(k-1)]^2,
\end{aligned}
\end{array}
\end{equation}
with zero initialized values $\hat{\sigma}_{wm}^2(0)=0$ for $m=1,...,M-1$, and the weighting factor $0<\gamma<1$. When the algorithm starts to converge, the estimation errors of the individual coefficients tend to become uncorrelated. On the other hand, when the unknown system abruptly changes, some of coefficients will encounter large variations. To track them, we propose to amend~\eqref{023} as
\begin{equation}
\label{023x2}
\begin{array}{rcl}
\begin{aligned}
\hat{\sigma}_{wm}^2(k) \leftarrow \max \left\lbrace \hat{\sigma}_{wm}^2(k),\hat{\sigma}_{w\cdot\text{avg}}^2(k)\right\rbrace,
\end{aligned}
\end{array}
\end{equation}
for $ m=1,...,M-1$, where
\begin{equation}
\label{023x3}
\begin{array}{rcl}
\begin{aligned}
\hat{\sigma}_{w\cdot\text{avg}}^2(k) = ||\bm w(k)-\bm w(k-1)||_2^2/M
\end{aligned}
\end{array}
\end{equation}
denotes the average uncertainty of all the coefficients.

3) In~\eqref{020}, we also note that another set of important parameters $\sigma_{\nu,i}^2$ that are the variances of subband noises. Inspired by~\cite{ni2010variable}, $\sigma_{\nu,i}^2$ can be effectively estimated as follows:
\begin{equation}
\label{024}
\begin{array}{rcl}
\begin{aligned}
\hat{\sigma}_{\nu,i}^2(k) = \hat{\sigma}_{e,i}^2(k) - \frac{||\hat{\bm r}_i(k)||_2^2}{\hat{\sigma}_{u,i}^2(k) + \epsilon_2},  \\
\end{aligned}
\end{array}
\end{equation}
where
\begin{equation}
\label{025}
\begin{array}{rcl}
\begin{aligned}
&\hat{\sigma}_{e,i}^2(k) = \beta \hat{\sigma}_{e,i}^2(k-1) + (1-\beta) q^2(e_{i,D}(k)) e_{i,D}^2(k),\\
&\hat{\sigma}_{u,i}^2(k) = \beta \hat{\sigma}_{u,i}^2(k-1) + (1-\beta) u_i^2(kN), \\
&\hat{\bm r}_i(k) = \beta \hat{\bm r}_i(k-1) + (1-\beta) q(e_{i,D}(k)) \bm u_i(k) e_{i,D}(k),\\
\end{aligned}
\end{array}
\end{equation}
with $\beta$ being the weighting factor which is chosen by $\beta=1/(\varrho M)$, $\varrho \geq 1$, and $\epsilon_2$ is a small positive number to avoid the division by zero. Note that, $\hat{\sigma}_{\nu,i}^2(k)$ in~\eqref{024} may be negative due to using the estimates from~\eqref{024}. To this end, we introduce the following limitation:
\begin{equation}
\label{025b}
\hat{\sigma}_{\nu,i}^2(k)=\left\{ \begin{aligned}
& \hat{\sigma}_{\nu,i}^2(k-1), \text{ if } \hat{\sigma}_{\nu,i}^2(k)  \leq 0\\
&\hat{\sigma}_{\nu,i}^2(k), \text{otherwise}.
\end{aligned} \right.
\end{equation}

As a result, Table~\ref{table_1} summarizes the proposed GR-SAF algorithm.
\begin{table}[tbp]
    \scriptsize
    \centering
    \vspace{-1em}
    \caption{Proposed GR-SAF Algorithm.}
    \label{table_1}
    \begin{tabular}{lc|}
        \hline\\
        \text{Initializations:} $\bm w(0) = \bm 0$,  $\bar{\bm \Phi}(0)=\frac{\epsilon_1}{M} \bm I_M$ with $\epsilon_1 \geq1$\\
        \text{Parameters:} $\epsilon_1 \geq1$, $\epsilon_2=10^{-5}$, $0\ll \gamma <1$, and $\beta=1/(\varrho M)$ with $\varrho \geq 1$\\
        \hline
        \textbf{For} \text{iteration} $k\geq 0$ in the decimated sequences\\
        \textbf{For} \text {each subband \emph{i}}\\
        \text{ }\text{ }\text{ }\text{ } $e_{i,D}(k)= d_{i,D}(k) - \bm u^\text{T}_i(k) \bm w(k-1)$\\
        \text{ }\text{ }\text{ }\text{ } $\left\{
        \begin{aligned}
        &\textbf{Calculation of the scaling parameters } q(e_{i,D}(k)):\\
        &\text{e.g., Eq.~\eqref{008} from the M-estimate or Eq.~\eqref{026a} from the MCC}\\
        \end{aligned} \right. $\\
        \text{ }\text{ }\text{ }\text{ } $\left\{
         \begin{aligned}
        &\textbf{Estimate of subband noises:} \\
        &\hat{\sigma}_{e,i}^2(k) = \beta \hat{\sigma}_{e,i}^2(k-1) + (1-\beta) q^2(e_{i,D}(k)) e_{i,D}^2(k)\\
        &\hat{\sigma}_{u,i}^2(k) = \beta \hat{\sigma}_{u,i}^2(k-1) + (1-\beta) u_i^2(kN) \\
        &\hat{\bm r}_i(k) = \beta \hat{\bm r}_i(k-1) + (1-\beta) q(e_{i,D}(k)) \bm u_i(k) e_{i,D}(k)\\
        &\hat{\sigma}_{\nu,i}^2(k) = \hat{\sigma}_{e,i}^2(k) - \frac{||\hat{\bm r}_i(k)||_2^2}{\hat{\sigma}_{u,i}^2(k) + \epsilon_2}\\
        &\hat{\sigma}_{\nu,i}^2(k)=\left\{ \begin{aligned}
        & \hat{\sigma}_{\nu,i}^2(k-1), \text{ if } \hat{\sigma}_{\nu,i}^2(k)  < 0\\
        &\hat{\sigma}_{\nu,i}^2(k), \text{otherwise}
        \end{aligned} \right.
        \end{aligned} \right. $\\

        \text{ }\text{ }\text{ }\text{ } $\bm g_i(k) = \frac{\bar{\bm \Phi}(k-1) \bm u_i(k) }{\bm u^\text{T}_i(k) \bar{\bm \Phi}(k-1) \bm u_i(k) + \bm u^\text{T}_i(k) \hat{\bm C}_w(k-1) \bm u_i(k) + \hat{\sigma}_{\nu,i}^2}$\\
        \text{ }\text{ }\text{ }\text{ } $\triangle_i(k) = q(e_{i,D}(k)) \bm g_i(k) e_{i,D}(k)$\\
        \text{ }\text{ }\text{ }\text{ } $\bm m_i(k) = \left[ 2q(e_{i,D}(k)) - q^2(e_{i,D}(k))\right] \bm g_i(k) \bm u^\text{T}_i(k)$ \\
        \text{ }\textbf{End}\\
        \text{ }$\bm w(k) = \bm w(k-1) + \sum \limits_{i=0}^{N-1} \triangle_i(k)$\\
        \text{ }$\hat{\sigma}_{wm}^2(k) = \gamma \hat{\sigma}_{wm}^2(k-1) + (1-\gamma)[w_m(k)-w_m(k-1)]^2$ \\
        \text{ }$\hat{\sigma}_{w\cdot\text{avg}}^2(k) = ||\bm w(k)-\bm w(k-1)||_2^2/M$ \\
        \text{ }$\hat{\sigma}_{wm}^2(k) = \max \left\lbrace \hat{\sigma}_{wm}^2(k),\hat{\sigma}_{w\cdot\text{avg}}^2(k)\right\rbrace,m=1,2,...,M$ \\

        \text{ }$\hat{\bm C}_w(k) = \text{diag}\{\hat{\sigma}_{w1}^2(k),\hat{\sigma}_{w2}^2(k),..., \hat{\sigma}_{wM}^2(k)\}$ \\
        \text{ }$\bar{\bm \Phi}(k) = \bar{\bm \Phi}(k-1) - \sum \limits_{i=0}^{N-1} \bm m_i(k) \bar{\bm \Phi}(k-1) + \hat{\bm C}_w(k)$ \\
        \text{ }\textbf{Setting the diagonal-off elements of $\bar{\bm \Phi}(k)$ to be zero.} \\
        \text{ }\textbf{End}\\
        \hline
    \end{tabular}
\end{table}

\section{Analysis and Discussion}
In this section, we discuss the generality, stability, convergence, and complexity of the GR-SAF algorithm.

\subsection{Generality of Algorithm}
As shown in Table~\ref{table_1}, we present a general framework for the robust SAF with the scaling parameter $q(e_{i,D}(k))$. Namely, by taking advantage of a different robust cost function~$\varphi(e_{i,D}(k))$, we can straightforwardly compute $q(e_{i,D}(k))$ by~\eqref{008} to arrive at a different robust GR-SAF algorithm. For example, based on the MCC strategy~\cite{chen2018mixture,chen2017kernel,chen2014steady}, the corresponding cost function is defined as $\varphi(e_{i,D}(k)) = \frac{1}{\sqrt{2\pi}\beta} \left[1- \exp \left( -\frac{e_{i,D}^2(k)}{2\kappa_\sigma^2}\right)\right]$, which leads to
\begin{equation}
\label{026a}
\begin{array}{rcl}
\begin{aligned}
q(e_{i,D}(k)) = \exp \left( -\frac{e_{i,D}^2(k)}{2\kappa_\sigma^2}\right),
\end{aligned}
\end{array}
\end{equation}
where $\kappa_\sigma>0$ is the kernel width. Similar to the aforementioned modified Huber criterion, when impulsive noise occurs, $q(e_{i,D}(k))$ resulted from the MCC criterion will be approaching 0 so that the MCC-based GR-SAF algorithm has good robustness against impulsive noise. Note that, the MCC-based SAF (MCC-SAF) algorithm was developed in~\cite{yu2016two}, but it required a trade-off between fast convergence and low steady-state error.

On the other hand, it is necessary to mention that the proposed GR-SAF algorithm also reduces to the general robust LMS (GR-LMS) algorithm when the number of subbands is one. The GR-LMS algorithm can also improves the existing robust LMS algorithms with specified robust criteria. However, their common problem is slow convergence for highly colored input signals, thus we do not discuss them in this paper.\\

In the following two subsections, we assume that the unknown system is time-invariant, i.e., $\bm c(k)=0$, to facilitate analysis.
\subsection{Stability of Algorithm}
Evidently, \eqref{022} describes the evolution behavior of the covariance matrix of the weight-error vector for the proposed algorithm. Thus, by taking the traces of all the terms in~\eqref{022}, the MSD recursion of the algorithm can be obtained:
\begin{equation}
\label{026}
\begin{array}{rcl}
\begin{aligned}
\text{MSD}(k) = \text{MSD}(k-1) - \sum \limits_{i=0}^{N-1} \rho_i(k) \bm u^\text{T}_i(k) \bar{\bm \Phi}(k-1) \bm g_i(k),
\end{aligned}
\end{array}
\end{equation}
which further yields
\begin{equation}
\label{027}
\begin{array}{rcl}
\begin{aligned}
\text{MSD}(k)& - \text{MSD}(k-1) \stackrel{(20)}{=} \\
& -\sum \limits_{i=0}^{N-1} \frac{ \rho_i(k) \bm u^\text{T}_i(k) \bar{\bm \Phi}^2(k-1) \bm u_i(k) }{\bm u^\text{T}_i(k) \bar{\bm \Phi}(k-1) \bm u_i(k) + \sigma_{\nu,i}^2}.
\end{aligned}
\end{array}
\end{equation}

Recalling that $\bar{\bm \Phi}(k)$ is a diagonal matrix with positive entries, denoted as $\bar{\bm \Phi}(k)=\text{diag}\{\sigma_{\Phi_1}^2(k),\sigma_{\Phi_2}^2(k),...,\sigma_{\Phi_M}^2(k)\}$ with $\sigma_{\Phi_m}^2(k)>0$, $m=1,2,...,M$, thus~\eqref{027} becomes
\begin{equation}
\label{027X}
\begin{array}{rcl}
\begin{aligned}
\text{MSD}(k)& - \text{MSD}(k-1) = \\
& -\sum \limits_{i=0}^{N-1} \frac{ \rho_i(k) \sum_{m=1}^M u^2_i(kN-m+1) \sigma_{\Phi_m}^4(k)}{\sum_{m=1}^M u^2_i(kN-m+1) \sigma_{\Phi_m}^2(k) + \sigma_{\nu,i}^2}.
\end{aligned}
\end{array}
\end{equation}
Moreover, there is $\rho_i(k)>0$ due to $0<q(e_{i,D}(k))<1$. It follows that the right side of~\eqref{027X} is always less than 0 so that~$\text{MSD}(k)-\text{MSD}(k-1)<0$, which ensures the convergence of the GR-SAF algorithm.
\subsection{Convergence of Algorithm}
It is further assumed that all the diagonal entries of $\bar{\bm \Phi}(k)$ are equal, namely, $\bar{\bm \Phi}(k)=\sigma_\Phi^2(k) \bm I_M$. Following this, both~\eqref{020} and~\eqref{022} are simplified respectively as
\begin{equation}
\label{028}
\begin{array}{rcl}
\begin{aligned}
\bm g_i(k) = \frac{\bm u_i(k) }{||\bm u_i(k)||_2^2 + \frac{\sigma_{\nu,i}^2}{\sigma_\Phi^2(k-1)}},
\end{aligned}
\end{array}
\end{equation}
and
\begin{equation}
\label{029}
\begin{array}{rcl}
\begin{aligned}
\sigma_\Phi^2(k) = \sigma_\Phi^2(k-1) - \sum \limits_{i=0}^{N-1} \rho_i(k) \frac{\sigma_{u,i}^2(k) \sigma_\Phi^2(k-1) }{||\bm u_i(k)||_2^2 + \frac{\sigma_{\nu,i}^2}{\sigma_\Phi^2(k-1)}}.
\end{aligned}
\end{array}
\end{equation}
By plugging \eqref{028} into \eqref{007}, then the weight update is formulated as
\begin{equation}
\label{030}
\begin{array}{rcl}
\begin{aligned}
\bm w(k) = \bm w(k-1) + \sum \limits_{i=0}^{N-1}  \frac{q(e_{i,D}(k)) \bm u_i(k) e_{i,D}(k)}{||\bm u_i(k)||_2^2 + \frac{\sigma_{\nu,i}^2}{\sigma_\Phi^2(k-1)}}.
\end{aligned}
\end{array}
\end{equation}
At a glance, in addition to having robustness against impulsive noise owing to the scaling parameter~$q(e_{i,D}(k))$, \eqref{030} behaves like an NSAF algorithm with the variable regularization parameter~$\delta(k)=\frac{\sigma_{\nu,i}^2}{\sigma_\Phi^2(k-1)}$. Indeed, it is. Thanks to the convergence of the proposed algorithm in Remark~2, $\sigma_\Phi^2(k)$ is large at the starting stage so that $\delta(k)$ is small, thereby speeding up the convergence. As the algorithm converges, $\sigma_\Phi^2(k)$ will gradually become very small; as such, $\delta(k)$ will becomes very large, finally achieving low steady-state estimation error.

\subsection{Complexity of Algorithm} Without considering~the computation of~the scaling parameters~$q(e_{i,D}(k))$ and the subband noise variances~$\sigma_{\nu,i}^2$, and recalling the diagonal simplification of $\bar{\bm \Phi}(k)$ in the previous practical consideration~1), the proposed GR-SAF algorithm requires $5MN+5M+2N^2+4N+1$ multiplications, $MN$ divisions, and $4MN+4M+2N^2+N-1$ additions per each iteration in the decimated sequences. As per each fullband input sample, this amount reduces to $5M+(5M+1)/N+2N+4+1$ multiplications, $M$ divisions, and $4M+(4M-1)/N+2N+1$ additions. In addition, the proposed algorithm has inherent computational complexity for partitioning both signals $u(n)$ and $d(n)$, which requires $2(J- 1)N$ additions and $2JN$ multiplications. Even so, the SAF is widely used in high-order applications such as the EC, thus $M$ would be much larger than the product $JN$; accordingly, the proposed algorithm still retain low computational complexity of $\mathcal{O}(M)$.

On the other hand, the data memory usage for the proposed GR-SAF algorithm when implementing it in embedded systems with limited memory, is $(5N+6)M+(2J+8)N+5$ in units of B-bit words. This amount is higher than that of the NSAF algorithm with $(N+2)M+(2J+3)N+3$, but is much less than that of the RLS algorithm with $2M^2+4M+5$~\cite{lee2009subband}.


 \section{Simulation results}
\begin{figure}[htb]
    \centering
    \includegraphics[scale=0.55] {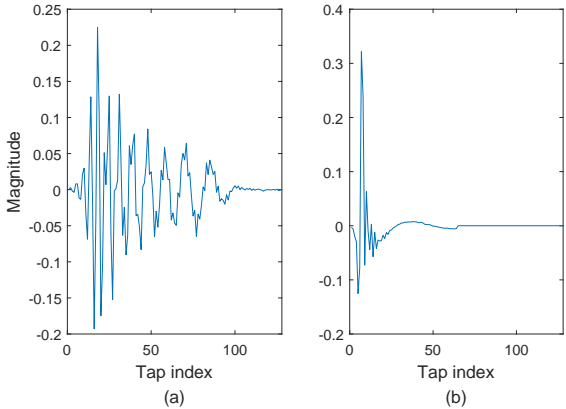}
    \hspace{2cm}\caption{Impulse responses of network echo channels: (a) dispersive channel, (b) sparse channel.}
    \label{Fig2}
\end{figure}

In this section, the proposed GR-SAF algorithm is evaluated in various examples. The length of the unknown system is assumed known and we set the length of the adaptive filters as the same of the unknown system. For the analysis filters~$\{H_i(z)\}_{i=0}^{N-1}$, we employ the cosine-modulated filter bank, and the coefficients of their impulse responses are expressed as
\begin{equation*}
\label{0sim1}
\begin{array}{rcl}
\begin{aligned}
h_i(l) = 2 p(l)\cos \left[ \frac{(2i+1)(2l-(L-1))\pi}{4N} + (-1)^i\frac{\pi}{4} \right],\\
\end{aligned}
\end{array}
\end{equation*}
where $p(l),\;l=0,1,...,J-1$ are the coefficients of the prototype filter. Herein, the lengths of the prototype filter for $N=2$, 4, and 8 subbands are $J=17$, 33, and 65, respectively, to obtain 60 dB stopband attenuation. The high stopband attenuation ensures almost no overlap between adjacent analysis filters as well as negligible cross-correlation between nonadjacent subbands. In our simulations, the unknown vectors $\bm w^o$ are the impulse responses of network echo channels shown in Fig.~\ref{Fig2}, i.e., channel-1 and channel-4 from the ITU-T G.168 standard~\cite{stnec2015} which are sparse and dispersive (non-sparse) respectively, with $M=128$ taps, unless otherwise specified. All the results are the average of 50 independent runs, except for the speech input scenario (single run).
\subsection{Scenario 1: System Identification}
The input signal $u(n)$ is generated from the first-order autoregressive (AR) system with a pole at 0.95, which has high correlation with the eigenvalue spread of $\chi=1337$ as compared to the white input signal of $\chi=1$. The additive noise $\nu(n)$ is a contaminated-Gaussian process that consists of the Gaussian noise $\nu_g(n)$ and the impulsive noise $\nu_{im}(n)$, i.e., $\nu(n)=\nu_{g}(n) + \nu_{im}(n)$. The variance $\sigma_\text{g}^2$ of the zero-mean Gaussian noise $\nu_g(n)$ is given by the signal-to-noise ratio (SNR) defined as $10\log10(\sigma_{\bar{d}}^2/\sigma_g^2)$, where $\sigma_{\bar{d}}^2=\text{E}\{(\bm u_n^\text{T} \bm w^o)^2\}$ is the power of the system's output signal. The impulsive noise is generated by the model~$\nu_{im}(n)=b(n) \eta(n)$, where $b(n)$ is a Bernoulli random process with the probabilities of 1 and 0 being $p_r$ and $1-p_r$ respectively, and $\eta(n)$ follows from the white Gaussian distribution with zero-mean and variance~$\sigma_\eta^2 =1000 \sigma_{\bar{d}}^2$. Obviously, $p_r$ also represents the probability of appearing impulsive noise samples, and generally its value is small like $p_r=0.001$. The MSD is used as the performance measure.

\begin{figure}[htb]
    \centering
    \includegraphics[scale=0.5] {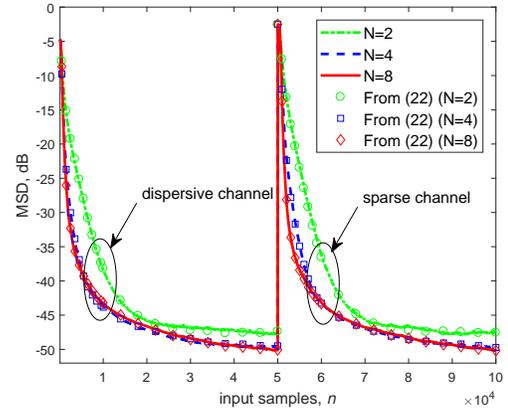}
    \hspace{2cm}\caption{MSD curves of the GR-SAF algorithm for different~$N$.}
    \label{Fig3}
\end{figure}

Fig.~\ref{Fig3} depicts the performance of the GR-SAF algorithm using $N=2$, 4, and 8. The M-estimate's parameters are set to $\tau=2$ and $N_w=20$. The GR-SAF's parameters are chosen as $\epsilon_1=1$, $\epsilon_2=10^{-5}$, $\gamma=0.95$ and $\varrho=2$. Clearly, by increasing $N$, the GR-SAF algorithm has better convergence, because larger $N$ lets the decimated subband inputs $\{\bm u_i(k)\}_{i=0}^N$ be closer to white. Naturally, this property of increasing $N$ will not be obvious after it is larger than a certain value such as 4 here, since at this moment the decimated subband inputs have been sufficiently whitened. As one can also see from Fig.~\ref{Fig3}, the GR-SAF algorithm using the diagonal approximation of $\bar{\Phi}(k)$ retains comparable performance with that using the full update of $\bar{\Phi}(k)$ given in~\eqref{022}.

\begin{figure}[htb]
    \centering
    \subfigure[]{
        \includegraphics[scale=0.5]{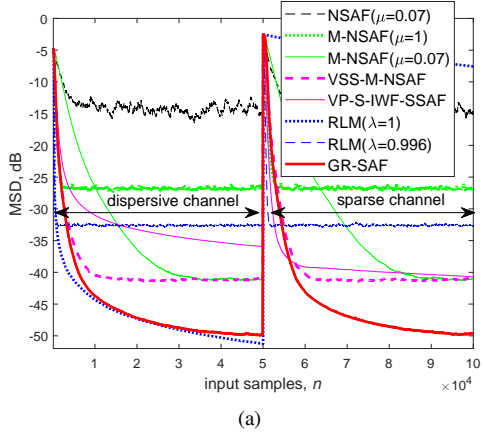}
        \label{Fig4a}
    }
    \subfigure[]{
        \includegraphics[scale=0.5]{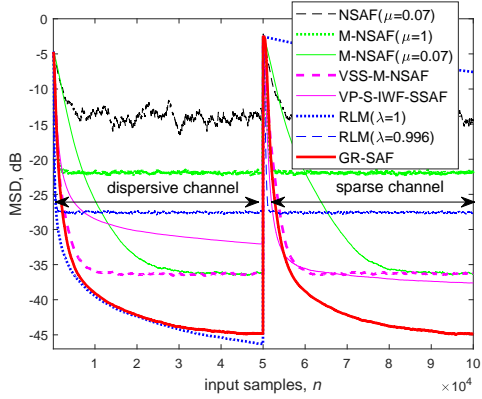}
        \label{Fig4b}
    }
    \caption{MSD curves of RLM and various subband algorithms with $N=4$. (a) SNR=30 dB, (b) SNR=25 dB.}
    \label{Fig4}
 \end{figure}

\begin{figure}[htb]
    \centering
    \subfigure[]{
        \includegraphics[scale=0.5]{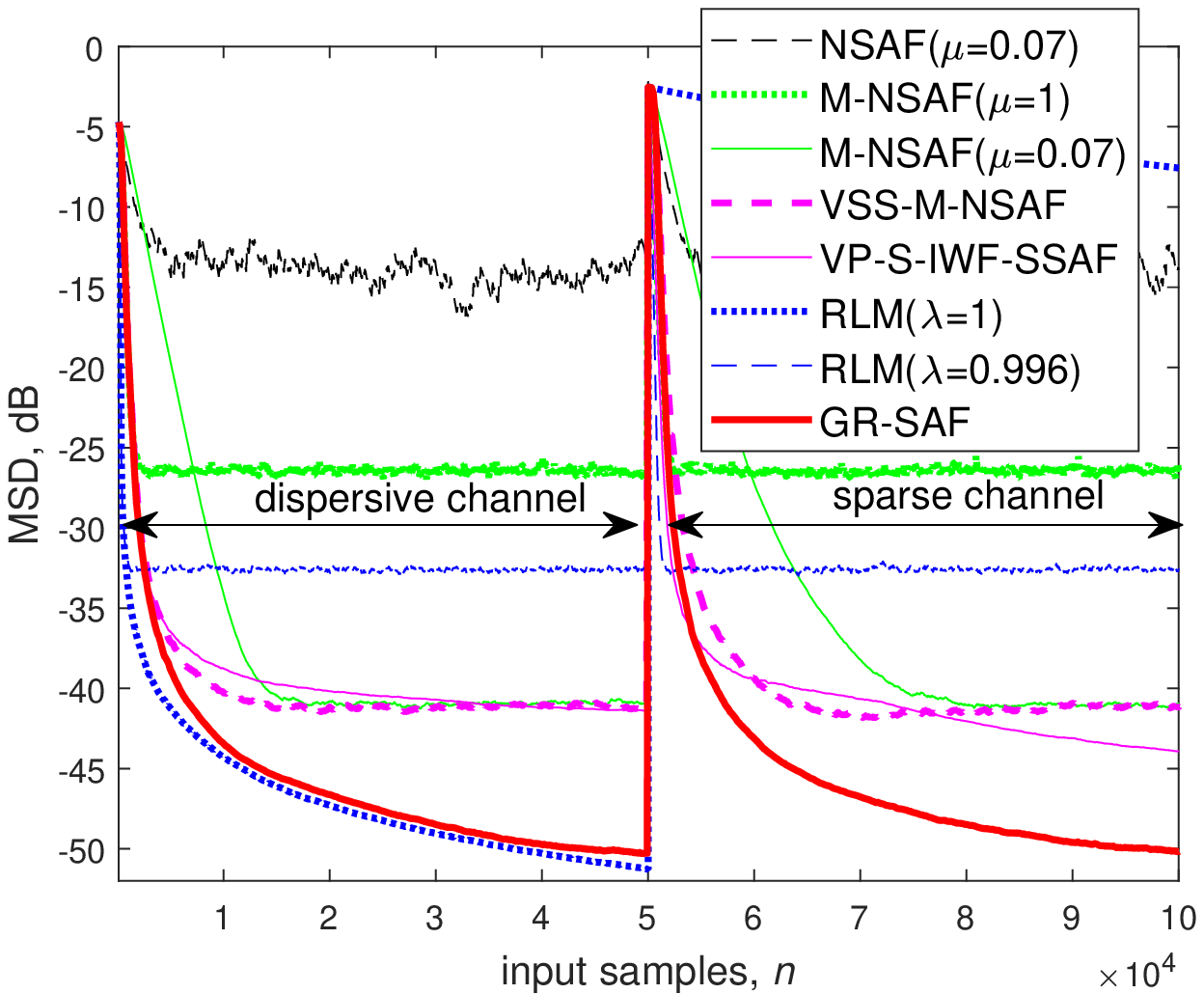}
        \label{Fig5a}
    }
    \subfigure[]{
        \includegraphics[scale=0.5]{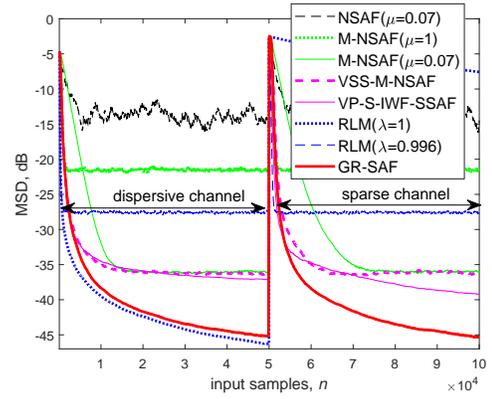}
        \label{Fig5b}
    }
    \caption{MSD curves of RLM and various subband algorithms with $N=8$. (a) SNR=30 dB, (b) SNR=25 dB.}
    \label{Fig5}
\end{figure}

In Figs.~\ref{Fig4} and \ref{Fig5}, we examine the performance of NSAF, VP-S-IWF-SSAF~\cite{YU2021107806}, and four M-estimate based algorithms including M-NSAF~\cite{yu2019m}, VSS-M-NSAF~\cite{yu2019m}, recursive least M-estimate (RLM)~\cite{chan2004recursive}, and proposed GR-SAF. To fairly compare them, the parameters of the subband-based algorithms are chosen according to the fact that they arrive at the almost same steady-state result or convergence rate. These M-estimate based algorithms use the MH function shown in~\eqref{005} whose parameters are the same as in Fig.~\ref{Fig3}. Other parameters of some algorithms are as follows~\footnote{For the compared algorithms, we use the same notations of parameters in references.}: $\mu_\text{min}=10^{-5}$, $\chi=1$, $\tau=1$, and $\xi=0.01$ for VP-S-IWF-SSAF; $\epsilon_1=10^{-6}$ and $\theta_\chi=5$ for VSS-M-NSAF; $\bm P(0)=(1/0.05)\bm I_M$ for RLM. As can be seen, the NSAF algorithm performs poorly in impulsive noise, while other algorithms behave with good robustness due to the suppression of the large magnitude influence of impulsive samples. As compared to the M-NSAF algorithm with the fixed step-size, both VSS-M-NSAF and GR-SAF algorithms obtain fast convergence and low steady-state MSD. It should be stressed that the VSS-M-NSAF algorithm employ the time-varying step-size scheme,  while the GR-SAF is like a variable regularization M-NSAF algorithm as stated in Remark~3 and achieves better performance than the former. The VP-S-IWF-SSAF algorithm exhibits better performance for identifying a sparse channel than a dispersive channel, owing to this fact that it takes advantage of the sparsity of the underlying systems. Among these algorithms, the convergence of the RLM algorithm with $\lambda=1$ is the best, because it has stronger decorrelation ability on input signals than the subband decomposition. However, the complexity of the RLM algorithm with the level of $\mathcal{O}(M^2)$ is much higher than that of the GR-SAF algorithm with the level of $\mathcal{O}(M)$, especially for large~$M$. In a nutshell, the proposed GR-SAF algorithm can approximate the RLM performance with $\lambda=1$ well, but with lower complexity. Furthermore, the RLM with $\lambda=1$ loses the re-convergence (or tracking) ability after the target vector shows a sudden change; conversely, using smaller $\lambda=0.996$ improves the tracking ability of the RLM algorithm, but also brings about large steady-state MSD.

\begin{figure}[htb]
    \centering
    \subfigure[]{
        \includegraphics[scale=0.46]{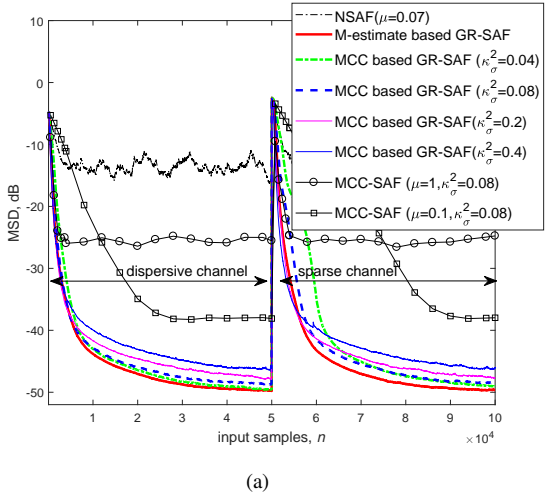}
        \label{Fig6a}
    }
    \subfigure[]{
        \includegraphics[scale=0.46]{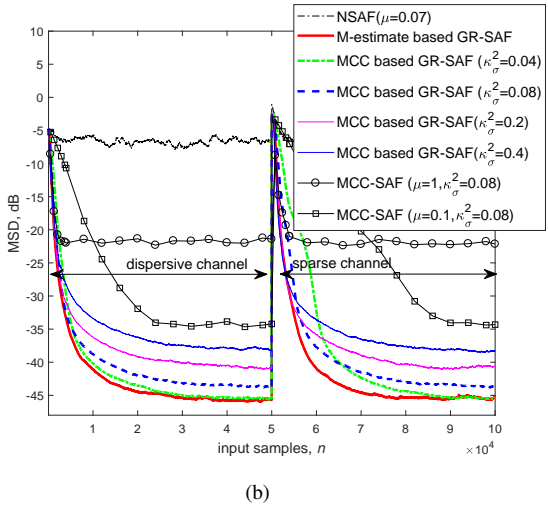}
        \label{Fig6b}
    }
    \caption{MSD performance of M-estimate and MCC based GR-SAF algorithms. (a) $p_r=0.001$, (b) $p_r=0.005$. [$N=4$, SNR=30 dB].}
    \label{Fig6}
\end{figure}

In Fig.~\ref{Fig6}, we investigate the performance of the GR-SAF update using different scaling factors $q(e_{i,D}(k))$ from the M-estimate (Eq.~\eqref{008}) and MCC (Eq.~\eqref{026a}) criteria, respectively. Clearly, when using the MCC criteria, the proposed GR-SAF framework significantly improves the MCC-SAF's performance~\cite{yu2016two} in terms of convergence rate and steady-state error. It means that the GR-SAF update has good performance, in this case whose performance depends only on the scaling factor $q(e_{i,D}(k))$ stemming from some specified robust criteria. For the MCC example, the kernel width $\kappa_\sigma$ controls the trade-off between the robust and convergence levels of the GR-SAF update. However, discussing the effects of different $q(e_{i,D}(k))$ is not the focus of this paper and in the sequel we only display the results of the GR-SAF algorithm based on the M-estimate.
\subsection{Scenario 2: Network Echo Cancellation}
\begin{figure}[htb]
    \centering
    \includegraphics[scale=0.5] {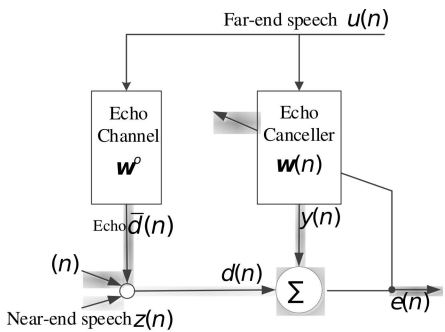}
    \hspace{2cm}\caption{Structure of an adaptive echo canceler.}
    \label{Fig1x}
\end{figure}

\begin{figure}[htb]
    \centering
    \includegraphics[scale=0.5] {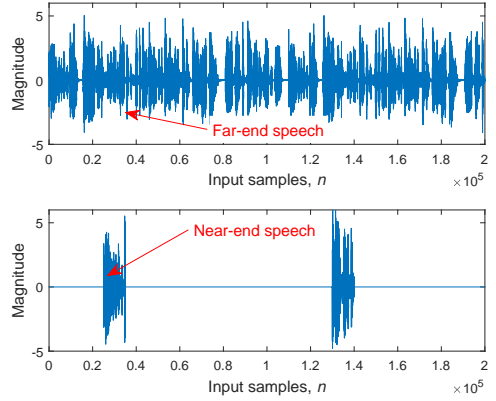}
    \hspace{2cm}\caption{Speech signals.}
    \label{Fig8}
\end{figure}

Echo canceler is vital for hands-free speech communication and teleconferencing systems. Fig.~\ref{Fig1x} shows the general structure of adaptive echo cancellation, which covers the network echo cancellation~\cite{paleologu2010efficient,yang2011proportionate} and acoustic echo cancellation scenarios~\cite{huang2021affine,zhou2011new}. In this scenario, the speech input signal $u(n)$ from the far-end talker's is convoluted with the echo channel and the resulting signal $\bar{d}(n)$ is the annoying echo, where the network echo channel is due to mixed packet-switched and circuit-switched components (e.g., Fig.~\ref{Fig2}), while the acoustic echo channel is due to the coupling between the loudspeaker and microphone (e.g., Fig.~\ref{Fig13}). Using the same input signal, the echo canceler first identifies the echo channel $\bm w^o$ with the adaptive filter~$\bm w(n)$, and the corresponding output signal $y(n)$ is the replica of the echo. Then, by subtracting $y(n)$ from the desired signal $d(n)$ that could consists of the echo $\bar{d}(n)$, the near-end speech $z(n)$, and the additive noise $\nu(n)$, the echo will be eliminated to improve the call quality. Compared with system identification, the main difference in echo cancellation is using the realistic speeches as the input. In this subsection, the network echo cancellation is considered, where the employed far-end and near-end speeches are shown in Fig.~\ref{Fig8}.

\begin{figure}[htb]
    \centering
    \includegraphics[scale=0.5] {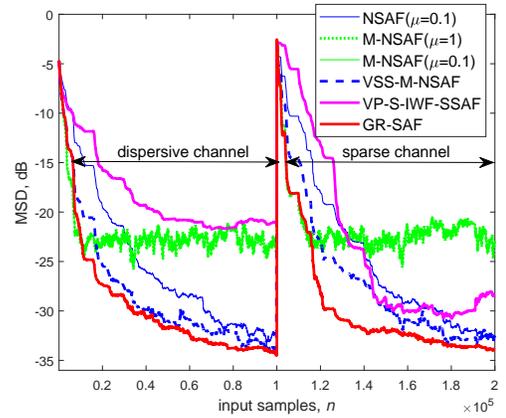}
    \hspace{2cm}\caption{MSD curves of various subband algorithms in the Gaussian noise. [$N=4$, SNR=30 dB]. Parameters of some algorithms are reset as follows: $\chi=3$ (VP-S-IWF-SSAF); $\gamma=0.99$ and $\varrho=3$ (GR-SAF).}
    \label{Fig9}
\end{figure}

\begin{figure}[htb]
    \centering
    \includegraphics[scale=0.5] {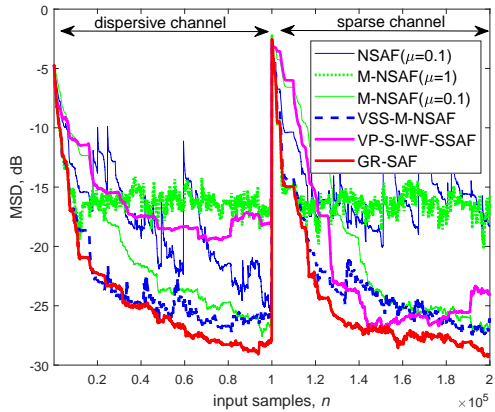}
    \hspace{2cm}\caption{MSD curves of various subband algorithms in the $\alpha$-stable noise. [$N=4$, $\alpha=1.6$ and $\vartheta=1/30$].}
    \label{Fig10}
\end{figure}

Case~1: there is no near-end speech, i.e., $z(n)=0$. Figs.~\ref{Fig9} and~\ref{Fig10} show the MSD performance of the above algorithms in the Gaussian and the $\alpha$-stable noise scenarios respectively. The characteristic function describing the symmetric~$\alpha$-stable distribution~\cite{nikias1995signal} is  $\phi(t)=\exp(-\vartheta |t|^\alpha)$, where the characteristic exponent $\alpha \in (0,2]$ determines the impulsiveness of the noise (for lower values of $\alpha$, the noise has more impulsive behavior), and $\vartheta>0$ represents the dispersion degree of the noise. Note that it will reduce to the Gaussian noise when $\alpha=2$. To avoid the division by zero in the weight update during the silent period of speech signals, we add a regularization parameter $\delta= 20\sigma_{u,i}^2/N$ for NSAF, M-NSAF, and VSS-M-NSAF algorithms and $\delta= \sqrt{20\sigma_{u,i}^2}$ for VP-S-IWF-SSAF algorithm. As one can see, the results with the speech input are similar to those with the AR input, that is, the proposed GR-SAF algorithm is not only insensitive to the impulsive noise (e.g., the $\alpha$-stable noise) but also superior to the other counterparts in terms of convergence rate and steady-state error.

Case~2: double-talk case that has the near-end voiced speech in the Gaussian noise\footnote{ We do not show the near-end unvoiced speech as it behaves like the Gaussian in Case~1.}. Unlike the random impulsive noises in the previous figures, the near-end voiced speech would be the long-lasting bursts of impulses. In this situation, robust NSAF algorithms (including the proposed algorithm) could not be valid since they are commonly designed to suppress random realizations of impulsive noises. To inhibit the divergence of the echo canceler during double-talk, equipping the double-talk detector (DTD) is a standard procedure, where the well-known Geigel DTD~\cite{duttweiler1978twelve} has been in commercial use for many   years. No matter when the double-talk is detected, the adaptation of the filter weights will be inhibited. Nevertheless, detection errors for the DTD would occur, and these result in large amounts of divergence of the non-robust algorithms (e.g., NSAF) even if they have equipped the DTD. A proven way for dealing with the double-talk is that uses the combination of robust algorithms and the DTD~\cite{gansler2000double,vega2008new}. In other words, robust algorithms can remedy the false of DTD resulting from undetected near-end voiced speech samples.

\begin{figure}[htb]
    \centering
    \includegraphics[scale=0.5] {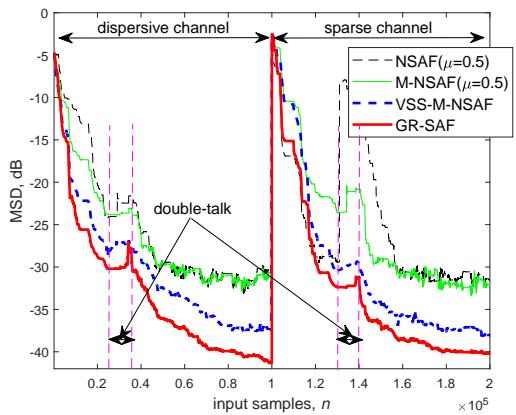}
    \hspace{2cm}\caption{MSD curves of various subband algorithms in a double-talk scenario. [$N=4$, SNR=35 dB]. In the double-talk, the M-estimate's parameters are readjusted as $N_w=40$ and $\theta=0.9995$. When using the Geigel DTD, we set $T_c=0.45$ and $T_{hold}=256$ in both NSAF and M-NSAF, and $T_c=0.45$ and $T_{hold}=40$ in both VSS-M-NSAF and GR-SAF.}
    \label{Fig11}
 \end{figure}

\begin{figure}[htb]
    \centering
    \includegraphics[scale=0.55] {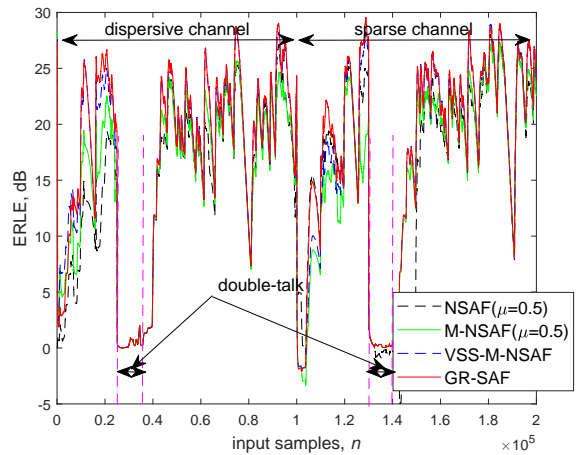}
    \hspace{2cm}\caption{ERLE curves of various subband algorithms in a double-talk scenario. [$N=4$, SNR=35 dB].}
    \label{Fig12}
 \end{figure}

For detecting the double-talk, the Geigel DTD is described as~\cite{duttweiler1978twelve}:
\begin{equation}
\label{036}
\begin{array}{rcl}
\begin{aligned}
d(kN) \geq T_c \max (|u(kN)|,|u(kN-1)|,...\\
|u(kN-M+1)|),
\end{aligned}
\end{array}
\end{equation}
where $0<T_c<1$ is a detector threshold. Once~\eqref{036} holds, the double-talk will be declared and meanwhile the adaptive filter will stop updating during the hangover period of $T_{hold}$. Similar to~\cite{yu2019m}, we also revise the median vector $\bm a_{e,i}$ in the M-estimator as $\bm a_{e,i}(k)=[r_i(k),r_i(k-1),...,r_i(k-N_w+1)]$, $i=0,...,N-1$, where $r_i(k) = 0$ when the double-talk is detected and during the period of $T_{hold}$, otherwise $r_i(k) = e_{i,D}^2(k)$. In addition to the MSD performance, the echo return loss enhancement (ERLE) is another performance measure, defined by $\text{ERLE}(n) = 10\log_{10} (\text{avg} \{d^2(n)\}/\text{avg} \{e^2(n)\})$~\cite{hansler2005acoustic}, where $\text{avg}(\cdot)$ is a smooth operator having the form $\text{avg} \{d^2(n)\}=0.999\text{avg} \{d^2(n)\} + 0.001d^2(n)$ with $\text{avg} \{d^2(0)\}=0$. In Figs.~\ref{Fig11} and~\ref{Fig12}, we compare the MSD and ERLE behaviors of the algorithms in a double-talk scenario, respectively. As can be seen from, under using the same step-size and the parameters of Geigel DTD, the M-NSAF algorithm is more robust to double-talk than the NSAF algorithm, due to the fact that the former has the capability of distinguishing outliers. Moreover, among these improvements of M-NSAF algorithm, the proposed GR-SAF algorithm achieves faster convergence and smaller steady-state MSD or higher steady-state ERLE as compared to the VSS-M-NSAF algorithm.
\subsection{Scenario 3: Acoustic Echo Cancellation}
\begin{figure}[htb]
    \centering
    \includegraphics[scale=0.55] {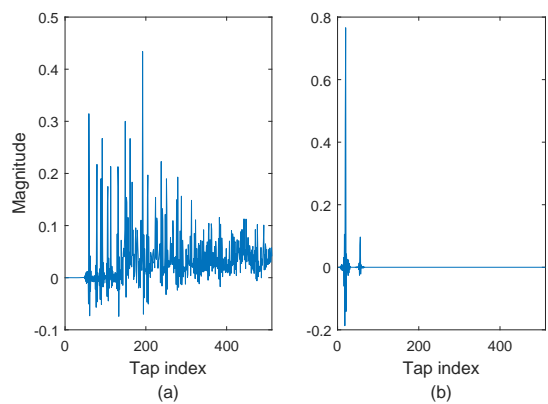}
    \hspace{2cm}\caption{Impulse responses of acoustic echo channels: (a) dispersive channel from~\cite{yu2016novel}, (b) sparse channel~\cite{YU2021107806}.}
    \label{Fig13}
\end{figure}

\begin{figure}[htb]
    \centering
        \includegraphics[scale=0.5]{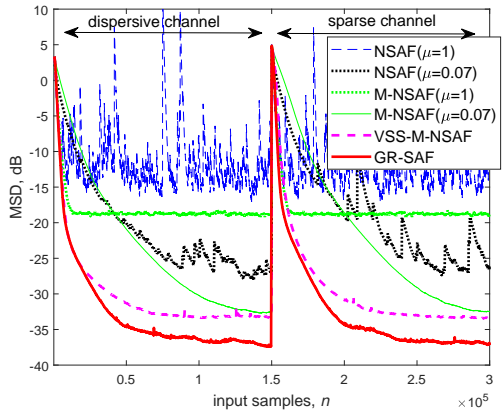}
    \caption{MSD curves of various subband algorithms in the $\alpha$-stable noise (with $\alpha=1.6$ and $\vartheta=1/30$). [$N=4$]. The M-estimate's parameters are set to $\tau=2$ and $N_w=30$. We set other parameters of algorithms as follows: $\epsilon_1=10^{-6}$ and $\theta_\chi=5$ for VSS-M-NSAF; $\gamma=0.99$ and $\varrho=3$ for GR-SAF.}
    \label{Fig14}
\end{figure}

In this subsection, we examine the proposed GR-SAF algorithm in acoustic echo cancellation scenarios. The impulse responses of acoustic echo channels truncated to 512 taps are plotted in Fig.~\ref{Fig13}. As one can see, due to the time delay difference among the reflections, the channel (b) is more sparser than the channel (a). It is well known that this difference depends on the characteristics of acoustic environments, for example, in either an indoor environment or an outdoor environment, temperature, pressure, and the volume of the enclosed space. Fig. \ref{Fig14} shows the MSD curves of the algorithms using an AR input. It is clear to see that the sparseness of acoustic echo channels does not affect the performance of the GR-SAF algorithm, and with working better than the existing algorithms in the $\alpha$-stable noise.

\begin{figure}[htb]
    \centering
    \includegraphics[scale=0.55] {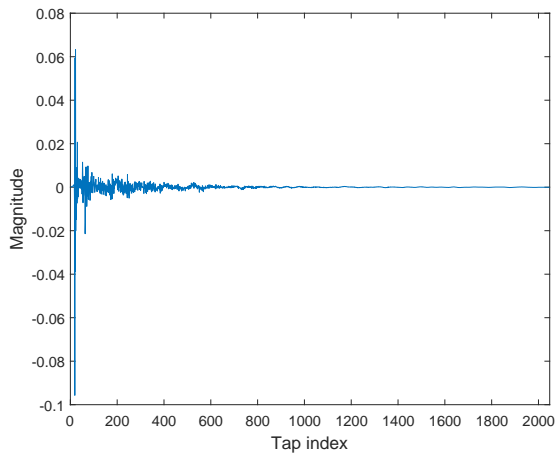}
    \hspace{2cm}\caption{A measured acoustic impulse response.}
    \label{Fig15}
\end{figure}

\begin{figure}[htb]
    \centering
    \includegraphics[scale=0.5]{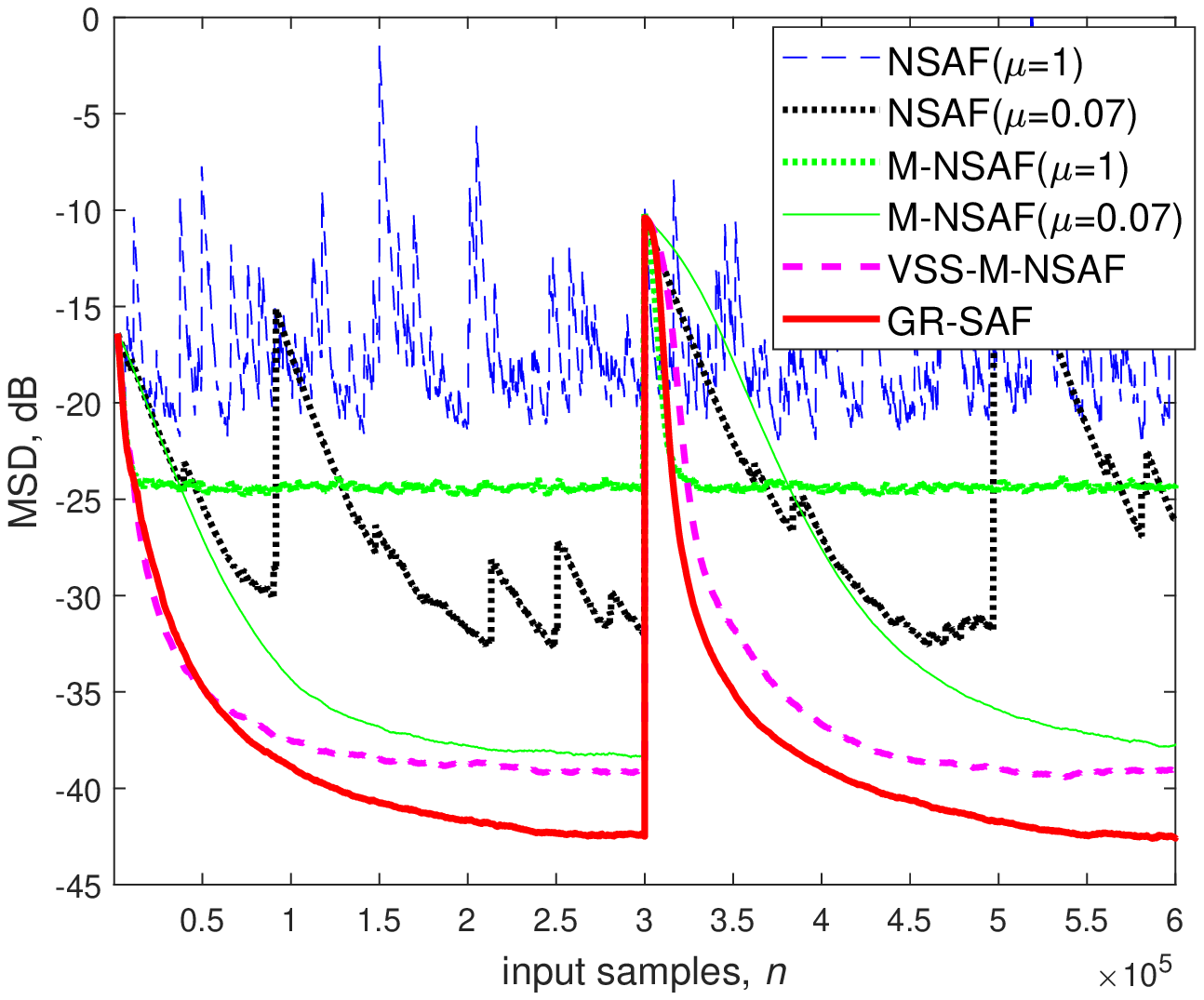}
    \caption{MSD curves of various subband algorithms in the $\alpha$-stable noise (with $\alpha=1.6$ and $\vartheta=1/50$) using an AR input. [$N=4$]. The M-estimate's parameters are set to $\tau=2$ and $N_w=30$. We set other parameters of algorithms as follows: $\epsilon_1=10^{-6}$ and $\theta_\chi=5$ for VSS-M-NSAF; $\gamma=0.99$ and $\varrho=2$ for GR-SAF.}
    \label{Fig16}
\end{figure}

\begin{figure}[htb]
    \centering
    \includegraphics[scale=0.55] {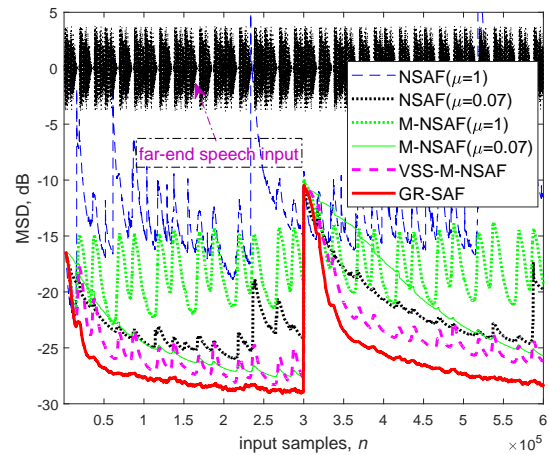}
    \hspace{2cm}\caption{MSD curves of various subband algorithms in the $\alpha$-stable noise (with $\alpha=1.6$ and $\vartheta=1/50$) using a speech input. [$N=4$]. Parameters of algorithms are chosen to be the same as Fig.~\ref{Fig16}.}
    \label{Fig17}
\end{figure}

On the other hand, a measured acoustic impulse response in a 6m $\times$ 6m $\times$ 2.4m room (the speech and acoustic lab of the Faculty of Engineering at BarIlan University (BIU)) is also considered~\cite{hadad2014multichannel}, with the reverberation time 160 ms. It is truncated to 2048 taps and shown in~\ref{Fig15}. In Figs.~\ref{Fig16} and~\ref{Fig17}, we evaluate the performance of the algorithms for the AR input and the speech input, respectively, where the acoustic impulse response is changed by multiplying $-1$ at the middle of input samples. It is seen from that the proposed GR-SAF algorithm still have better performance than its competitors.

\section{Conclusion}
Based on the random-walk model with individual coefficient uncertainty, we minimize the MSD behavior to develop the GR-SAF algorithm's framework with robustness against impulsive noise. Accordingly, by designing different scaling factors such as from the M-estimate and maximum correntropy robust criteria, the GR-SAF algorithm can be obtained easily. On the other hand, the proposed GR-SAF algorithm can be simplified to be a robust NSAF algorithm, but with the time-varying regularization parameter, thus providing better performance in both convergence and steady-state behaviors. Simulation results in various scenarios have demonstrated the effectiveness of the proposed GR-SAF algorithm.

\ifCLASSOPTIONcaptionsoff
  \newpage
\fi

\bibliographystyle{IEEEtran}
\bibliography{IEEEabrv,mybibfile}

\end{document}